\documentclass[aps,preprint,showpacs,superscriptaddress,groupedaddress]{revtex4}  
\usepackage{graphicx}  
\usepackage{dcolumn}   
\usepackage{bm}        
\usepackage{amssymb}   
\begin{document}



\title{Effects of Mass Varying Neutrinos on Cosmological Parameters as determined from the Cosmic Microwave Background}
\author{Akshay Ghalsasi and Ann E. Nelson}
\email{aghalsa2@uw.edu, aenelson@uw.edu}
\affiliation{ Department of Physics, University of Washington, Seattle, WA 98195-1560}

\begin{abstract}
In models with a light scalar field (the `acceleron') coupled to neutrinos, neutrino masses depend on   neutrino density. The 
resulting coupled system of mass varying neutrinos (MaVaNs) and the acceleron can act as a negative pressure fluid and  is a candidate for  dark energy \cite{Fardon:2003eh} .   MaVaNs also allow for higher $\Sigma$m$_\nu$ than terrestrial bounds, giving late forming warm dark matter. In this paper we  study the effect of increasing neutrino mass on the CMB spectrum, implementing MaVaNs cosmology using CMBEASY.   We find that the CMB spectrum is only affected at very low multipoles. Cosmic variance allows for significant warm dark matter at late times. This implies that in MaVaNs cosmology  $\sigma_8$ as determined by the CMB  may not be a good determinant of structure evolution at late times, potentially reducing the tension between $\sigma$$_{8}$ as reported by Planck Collaboration  \cite{PlanckSZ:2013} without increasing the tension in  the Planck determined value of  Hubble's constant. In addition, in MaVaNs theories, CMB data do not necessarily constrain possible neutrino mass results in terrestrial experiments.
\end{abstract}

\maketitle

\section{Introduction}
The standard model of cosmology ($\Lambda$CDM) describes  cosmological data rather  well. In $\Lambda$CDM  the energy density of the universe has five components: cold dark matter, baryons, photons, neutrinos, and dark energy . Although neutrinos are assumed to be very light in $\Lambda$CDM it is a fairly straightforward extension to consider the case of more massive neutrinos. Since $\Lambda$CDM is such a good fit to the observed data, it is hard to concoct a model that changes significantly the evolution of the components of the total energy density at or before recombination. An alternative is to change the behavior of the components after recombination, preferably at late times, so that the CMB spectrum and distance to surface of last scattering  are not significantly affected.

The fact that at present times the dark energy density is of the same order of that of dark matter ($\rho$$_{CDM}$/$\rho$$_\Lambda$ = 1/3) is called the `cosmic coincidence problem'. We also have other `coincidences' in the fact that the other components of energy density i.e. baryons, photons, and neutrinos were also comparable to dark energy within a redshift of a few z.  This problem is puzzling because all the different components of the energy density redshift differently. The behavior of dark energy is not known at very high redshifts, but we know it redshifts very slowly at from z = 1 to present times ($\omega$ $\approx$ -1). If we assume that dark energy really is the cosmological constant i.e. $\omega$ = -1 throughout history then we have the coincidence problem mentioned above. On the other hand we can make dark energy `track' one of the other components such as dark matter or baryons until recent times after which dark energy switches to redshifting very slowly. Dark matter and baryons are consistent with redshifting as 1/a$^3$ since recombination. We can track baryons even further back to BBN. Hence it will be hard to postulate a model where dark energy tracks dark matter or baryons until recent times because that will only replace the coincidence problem with the `why now' problem, i.e. why did the dark energy stop tracking these components only recently.
Neutrinos offer more possibilities as the present energy density of neutrinos is not measured. We know that three species of neutrinos were relativistic at BBN until fairly recently, since non relativistic neutrinos act as hot dark matter and  tend to erase  structure on smaller scales.  
In ref. \cite{Fardon:2003eh} it was proposed that dark energy density tracks neutrino energy density, assuming that neutrino mass is not a fixed quantity but rather a dynamical one. If neutrinos couple to a light scalar field scalar field called the acceleron, the scalar field    effective potential is a function of the neutrino number density. For a broad range of acceleron potentials the acceleron evolves adiabatically, tracking the minimum of its effective potential, with the neutrino mass also evolving.  Depending on the form of the acceleron potential,  the neutrino-acceleron fluid together may produce   `dark energy' which can  explain the observed acceleration of the universe. Since the effective potential and hence the neutrino mass are a function of the number density, the model is called mass varying neutrinos or MaVaNs.

The aim of the present paper is to study the impact of MaVaNs on cosmology, particularly the CMB spectrum.
This will help us study the viability of MaVaNs as a theory and constrain its parameter space.

The next section discusses MaVaNs, their properties and some relations pertaining to their behavior. For more details and derivations consult \cite{Fardon:2003eh}.
\section{MaVaNs}
Mass Varying Neutrinos are neutrinos whose mass varies as a function of their number density and hence as a function of the scale factor. In one implementation of the theory the SM neutrinos get their mass from coupling to a sterile neutrino, which in turn gets its mass from the vev of a scalar field the `acceleron'. As the universe expands the sterile neutrino gets lighter and the SM neutrinos get heavier due to the to the see-saw mechanism. It can be shown \cite{Fardon:2003eh} that when the neutrinos are non relativistic the effective potential for the MaVaNs-acceleron fluid is given by
\begin{equation}
V(m_\nu) = m_\nu n_\nu + V_0(m_\nu)
\end{equation}
where $V_0$ is the acceleron potential. The minimum of this potential is given by 
\begin{equation}
V^{'}(m_\nu) =  n_\nu + V^{'}_{0}(m_\nu)
\end{equation}

The equation of state is then given by
\begin{equation}
\omega + 1 = -\frac{\delta log V}{3\delta log a}
           = \frac{m_\nu n_\nu}{V}
           = \frac{-m_\nu V_{0}^{'}(m_\nu)}{V(m_\nu)}
\end{equation}
where we have used equation [2] to get the second and third equality.
If we assume a power law dependence for the acceleron potential as a function of mass with a small exponent ($V_0$($m_\nu$) $\propto$ $m_\nu$$^{-k}$) then we get
\begin{equation}
\omega = \frac{-1}{1+k}
\end{equation}
Assuming that $\omega$ scales slowly with the scale factor, using equation [2] we can show that when the neutrinos are non relativistic 
\begin{equation}
m_\nu \propto \it a^{-3\omega}  
\end{equation}
We can also show that when the neutrinos are relativistic their mass scales as follows
\begin{equation}
m_\nu \propto \it a^{-(3\omega+1)/2}
\end{equation}
If $\omega$ = -1 as in the case of a cosmological constant then the SM neutrinos have a mass inversely proportional to their number density when non relativistic.

A potential problem with this theory is that when neutrinos become nonrelativistic an instability may develop where  the neutrinos clump on small scales, with the inter clump distance being large compared with the acceleron Compton wavelength, in which case the acceleron field is no longer smooth and no longer acts as dark energy \cite{matias}. We assume that such an instability does not develop, which can be arranged either  by  an acceleron potential which is sufficiently flat, or  a semi relativistic neutrino coupled to the acceleron \cite{Fardon:2006}.

Thus in this theory the neutrino mass becomes important only during recent times i.e. z $\approx$ few, depending on the present mass of the neutrino. Even if the neutrino mass today is several eV,  the neutrinos are nearly massless during recombination  and the mass does not directly affect the CMB spectrum. Indirectly, however, if $h$ is held constant, the MaVaNs theory would have a  different distance to the  last scattering   which would change the position of the acoustic peaks. It is therefore necessary to refit the cosmological parameters in the MaVaNs theory in order to obtain accord with the CMB fluctuation spectrum.

In figure 1  we have plotted comparisons of $\rho{_\nu}$ for $\Lambda$CDM and MaVaNs for different neutrino masses to compare how the neutrino energy density should vary in the two theories. In figure 2 we have plotted how neutrinos with the same mass but different $\omega$ vary as a function of the scale factor in the MaVaNs theory. It might seem that we are violating the terrestrial bounds on $\Sigma$m$_\nu$. However studies of gravitational clustering of massive neutrinos in the background of dark matter halos find significant overdensities can occur, thus possibly reducing the neutrino mass as measured by terrestrial experiments \cite{Singh:2002de} relative to the mass influencing larger scale cosmology.

Note that the sterile neutrinos introduced in the MaVaNs theory do not affect cosmology as they are much heavier  and out of thermal equilibrium at high redshift. Terrestrial evidence for sterile neutrinos with properties which would otherwise be incompatible with cosmology would be evidence in favor of MaVaNs\cite{Weiner:2005ac}.

\begin{figure}
\includegraphics[scale=1]{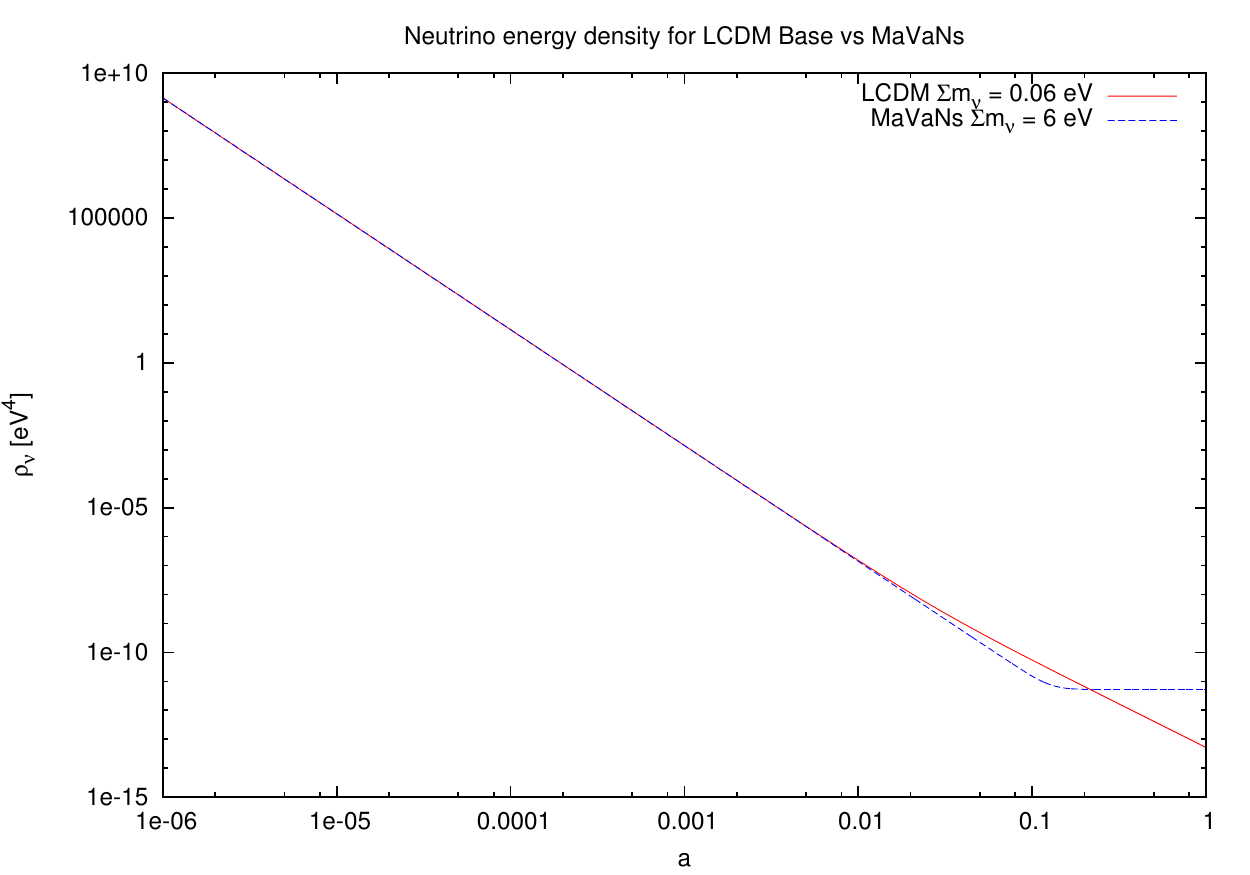}
\caption{\label{fig:epsart} Evolution of energy density for neutrinos in $\Lambda$CDM as compared to MaVaNs with $\omega$ = -1. `a'$=1/(1+z)$ is the scale factor.}
\end{figure}

\begin{figure}
\includegraphics[scale=1]{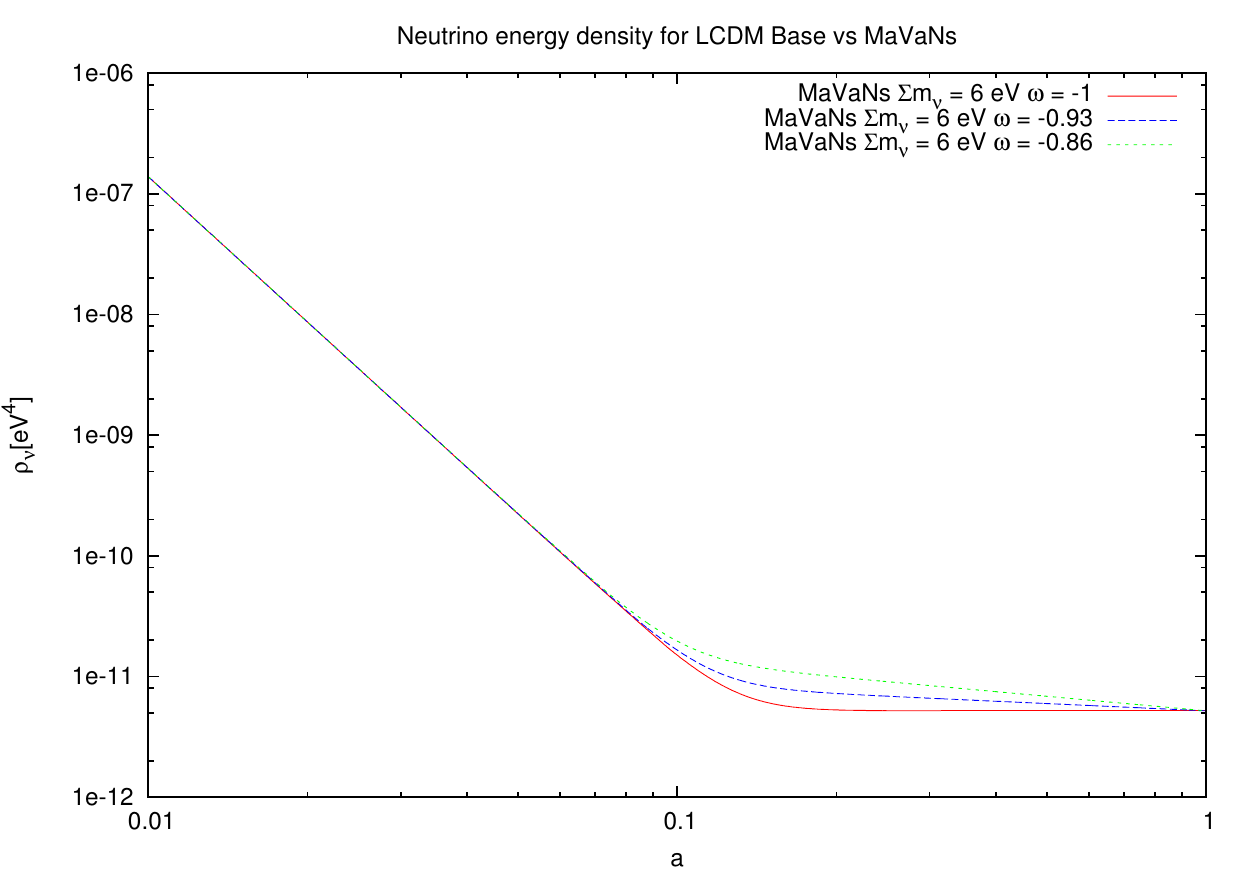}
\caption{\label{fig:epsart} Evolution of energy density for MaVaNs neutrinos with $\Sigma$m$_\nu$ = 6 eV but with  $\omega$ = -1, -0.93, -0.86. }
\end{figure}

Going back to equation [3] we can see for $\omega$ $\neq$ -1 the dark energy can also vary as a function of the scale factor. We find the quintessence energy density goes as follows in the non relativistic neutrino regime
\begin{equation}
V_{0}(m_\nu) \propto m_\nu^{-k} \propto a^{-\frac{3 k}{1+k}}
\end{equation}

As can be seen for $\omega$ = -1 i.e $\it k$ = 0 $V_0$ is a constant. But for $\omega$ $\neq$ -1 we get the acceleron potential to depend on the scale factor giving rise to varying dark energy density.

In our implementation of the MaVaNs neutrinos, the neutrinos act like matter after becoming non relativistic and are still influencing the evolution of dark energy.

\section{Implemenation}
We used the Planck Likelihood calculator for the range 50 $\le$ $\it l$ $\le$ 2500  to find our base $\Lambda$CDM model which was implemented using the publicly available code CMBEASY. We used a Metropolis algorithm to vary $\Omega_m$h$^2$, $\Omega_b$h$^2$, $\it h$ and the scalar amplitude $A_s$ to find our best fit parameters. We however kept $\tau$ = 0.0925 and $n_s$ = 0.9624 which are the best fit values obtained by Planck \cite{PlanckCosmoParameters:2013}. We assumed a flat universe and $\Omega_\Lambda$ was set such that \ $\Omega_{total}$ = 1. We also have $\Sigma$m$_\nu$ = 0.06 eV and N$_{eff}$ = 3.046. For our best fit values for the base model, we obtained $\Omega_m$h$^2$ = 0.1385, $\Omega_b$h$^2$ = 0.02197, $\it h$ = 0.686 and $\sigma_8$ = 0.8237. These are somewhat different from the best fit values obtained by Planck \cite{PlanckCosmoParameters:2013}. We attribute this to the fact that we are using a different code to calculate the CMB anisotropy (CMBEASY as opposed to CAMB) and possibly somewhat different nuisance parameters (see the Appendix for the list of nuisance parameters). The fact that our best fit model is somewhat different from Planck's best fit model does not affect the main point of the study, which is trying to compare different MaVaNs cosmologies with a base $\Lambda$CDM cosmology.  

We implemented MaVaNs cosmology by making modifications to CMBEASY. Different MaVaNs cosmologies are parametrized by having different $\omega$ and $\Sigma$m$_\nu$. We found the best fit values for each of the MaVans cosmologies by following the same likelihood minimization procedure above.

For the cases where $\omega$$ \neq$ -1 we use the Quintessence class in CMBEASY. Since CMBEASY dosen't have a class that models the MaVaNs potential, the specific potential we use is the inverse power law and we tune the exponent of our power law to give us dark energy density that we would expect from MaVaNs. Although we cannot get an exact MaVaNs like behavior for the quintessence energy density, we demand that they have similar values from z = 0 to z $\approx$ 4 after which the Dark Energy is subdominant compared to matter. Figures 3 and 4 illustrate the fact that the inverse power law is a good approximate fit to what we would expect from MaVaNs.

\begin{figure}                                                                                                                      
\includegraphics[scale = 1]{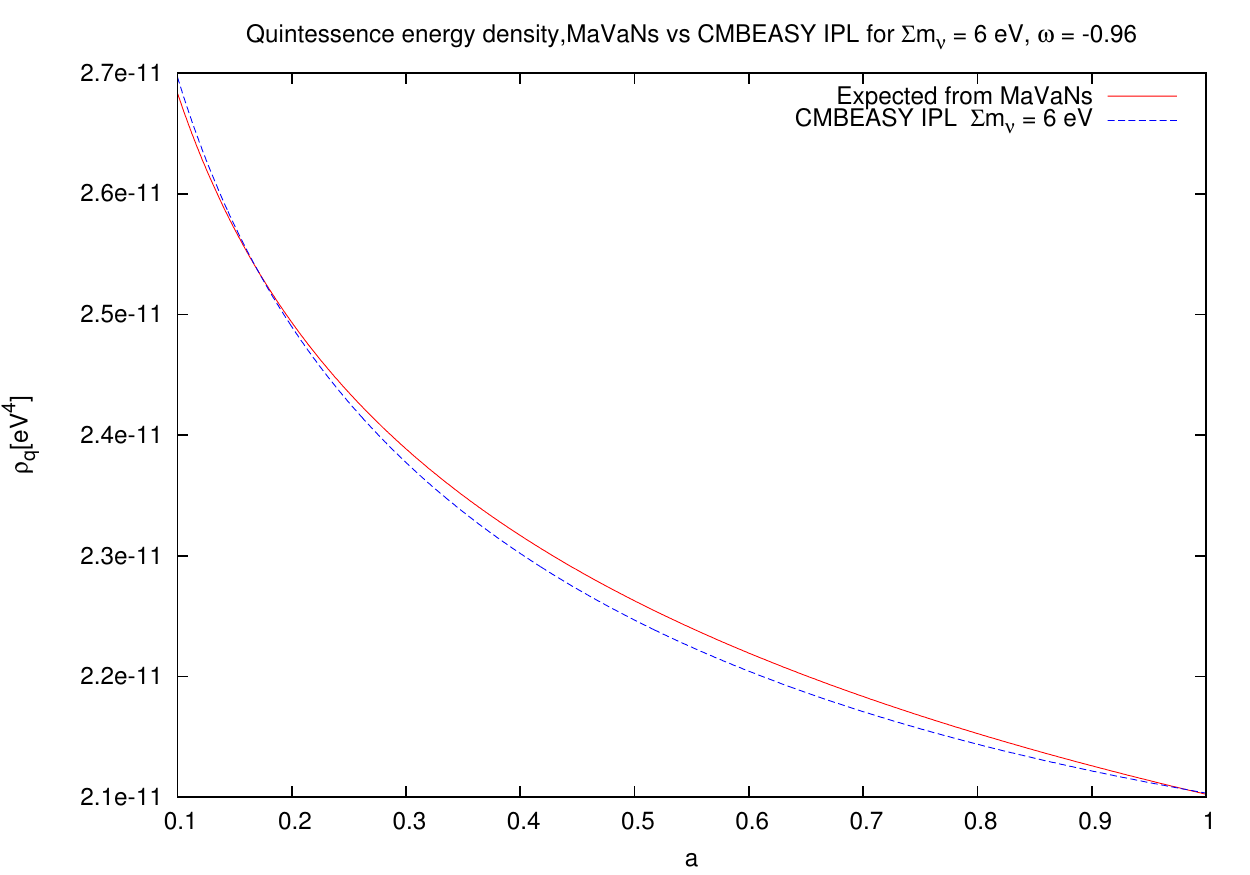}                                                                 
\caption{\label{fig:epsart} Expected quintessence energy density from MaVaNs for $\omega$ = -0.96 to that obtained from the inverse power law potential in CMBEASY. The exponent of the power law has been tuned such that the energy densities are very similar until about $a=1/(1+z)=0.2$. }
\end{figure}                 
                                                                                                
\begin{figure}                                                                                                                       
\includegraphics[scale = 1]{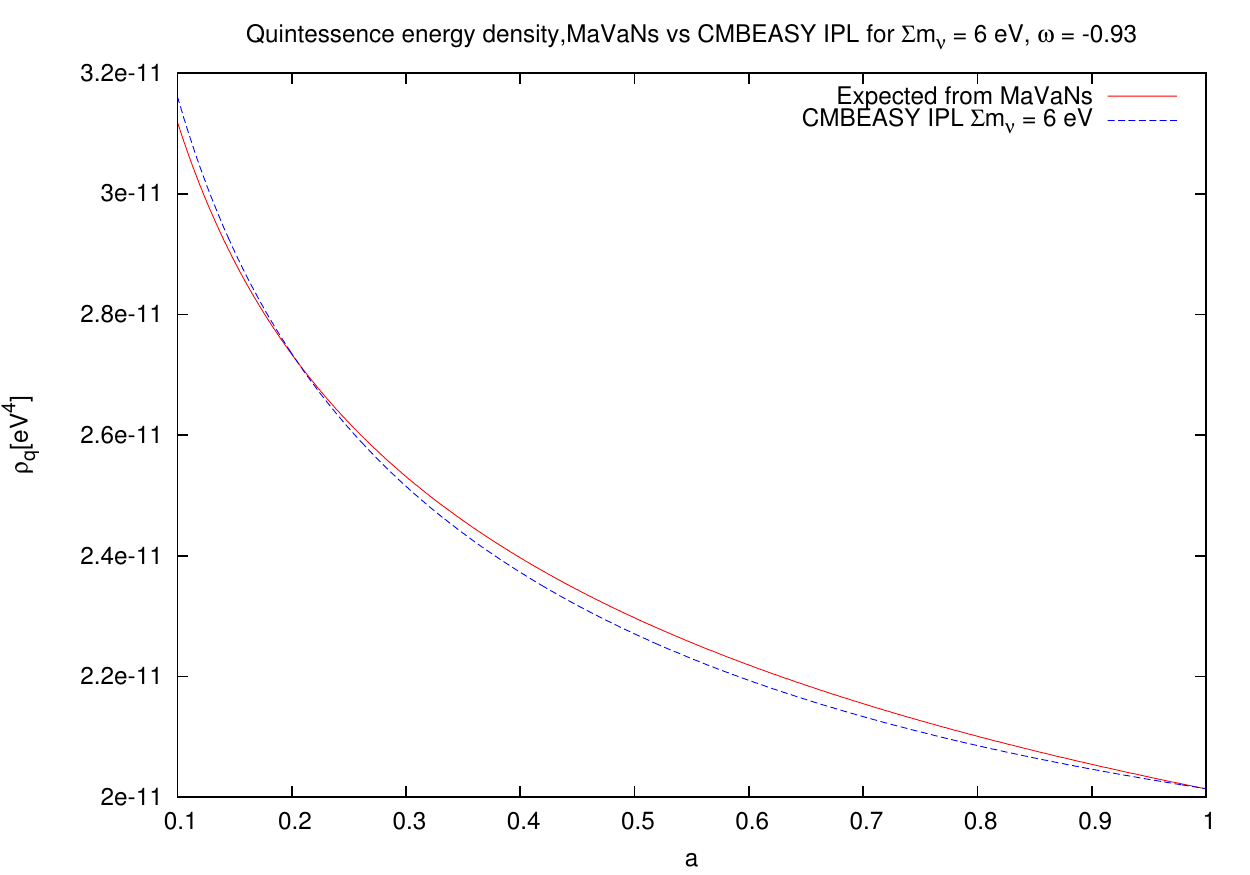}                                                               
\caption{\label{fig:epsart} Expected quintessence energy density from MaVaNs for $\omega$ = -0.93 to that obtained from the inverse power law potential in CMBEASY. The exponent of the power law has been tuned such that the energy densities are very similar until about  $a=1/(1+z)=0.2$. }                                                    
\end{figure} 

                                                                                         
\section{Analysis}

To compare the temperature power spectrum of a best fit MaVaNs model to the base $\Lambda$CDM model consult figure 5. As can be clearly seen the two spectra agree everywhere except at very low $\it l$, where the base MaVaNs spectrum gives a much larger $D_l$ than the base $\Lambda$CDM spectrum. The low $\it l$ spectrum is plotted in figure 6.

The low $\it l$  modes are affected by the late time ISW effect, which is increased because  we have  neutrinos which are acting like a significant amount of warm dark matter,  and therefore less dark energy. We can potentially use this late time ISW effect to put bounds on the current neutrino mass. However we first have to account for the cosmic variance.The cosmic variance of the quadrupole is given by $\Delta$$D_2\ $ = 0.63$D_2$. The error bars which are mostly due to cosmic variance have been plotted in figure 6. As can be seen both the $\Lambda$CDM and MaVaNs with $\Sigma$m$_\nu$ = 6 lie outside the 1$\sigma$ error bars. Although it is true that the discrepancy is slightly pronounced for MaVaNs with $\Sigma$m$_\nu$ = 6 as compared to $\Lambda$CDM, it still does not contribute significantly more to $\Delta$$\chi^2$. Moreover, the mechanism that is causing the low $\it l$ anomaly for $\Lambda$CDM, such as running of the spectral tilt, will affect MaVaNs as well making the MaVaNs spectrum for low $\it l$ and in better concordance with observations. For these reasons we consider $\Sigma$m$_\nu$ = 6 to be an acceptable present day neutrino mass.\footnote{Here we have only considered the quadrupole since looking at Planck data the effects of the ISW effect will be most pronounced for $\it l$ = 2. In practice we should consider all multipoles with $\it l$ $\le$ 50 that we have so far neglected in our likelihood calculations. These have been left for future studies and in principle will help put an upper bound on MaVaNs masses.}

\begin{figure}
\includegraphics[scale=1]{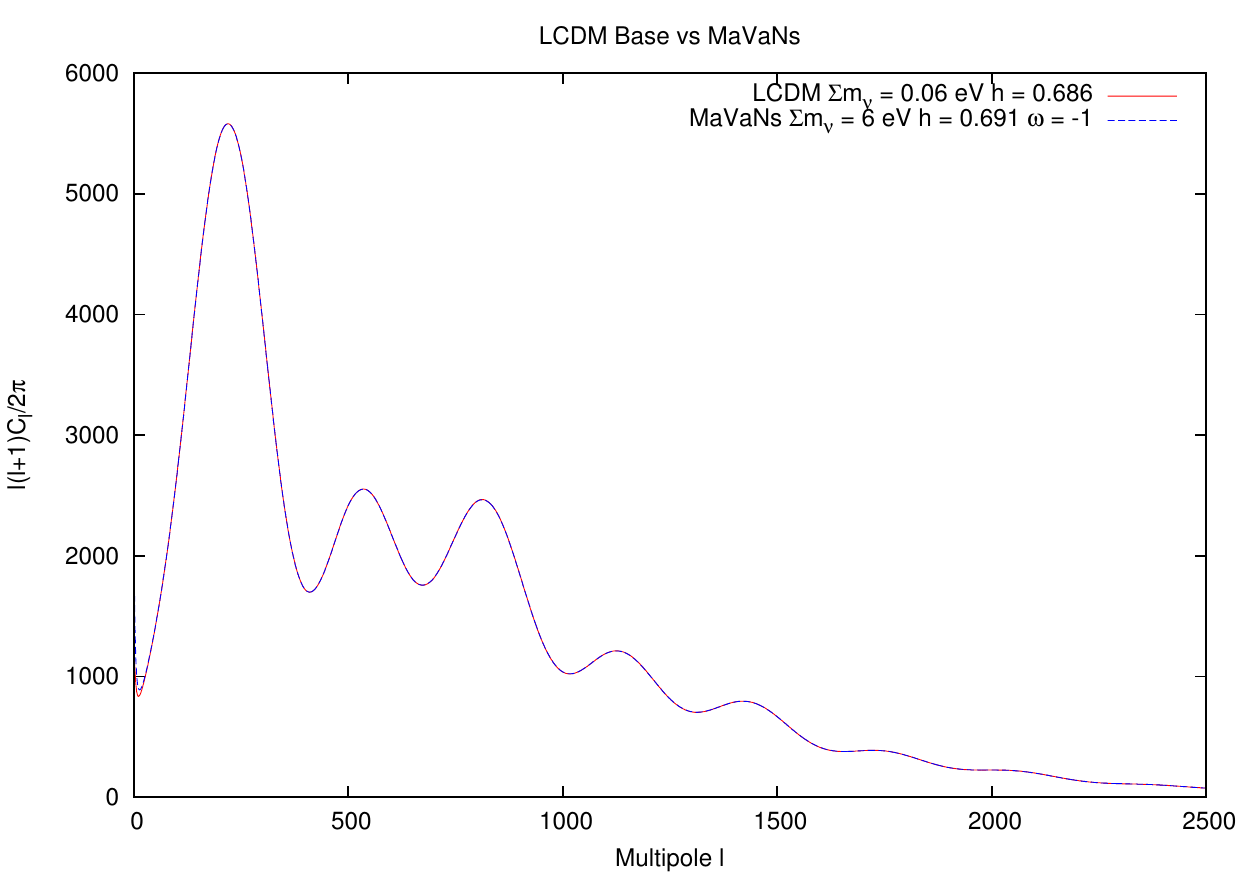}
\caption{\label{fig:epsart} The red curve shows the best fit temperature spectrum for our base $\Lambda$CDM cosmology.The blue curve show the best fit temperature spectrum for a MaVaNs cosmology with $\omega$ = -1 and $\Sigma$m$_\nu$ = 6 eV.As we can see the two curves agree really well for most values of $\it l$ except for the very low $\it l$. }
\end{figure}

\begin{figure}
\includegraphics[scale=1]{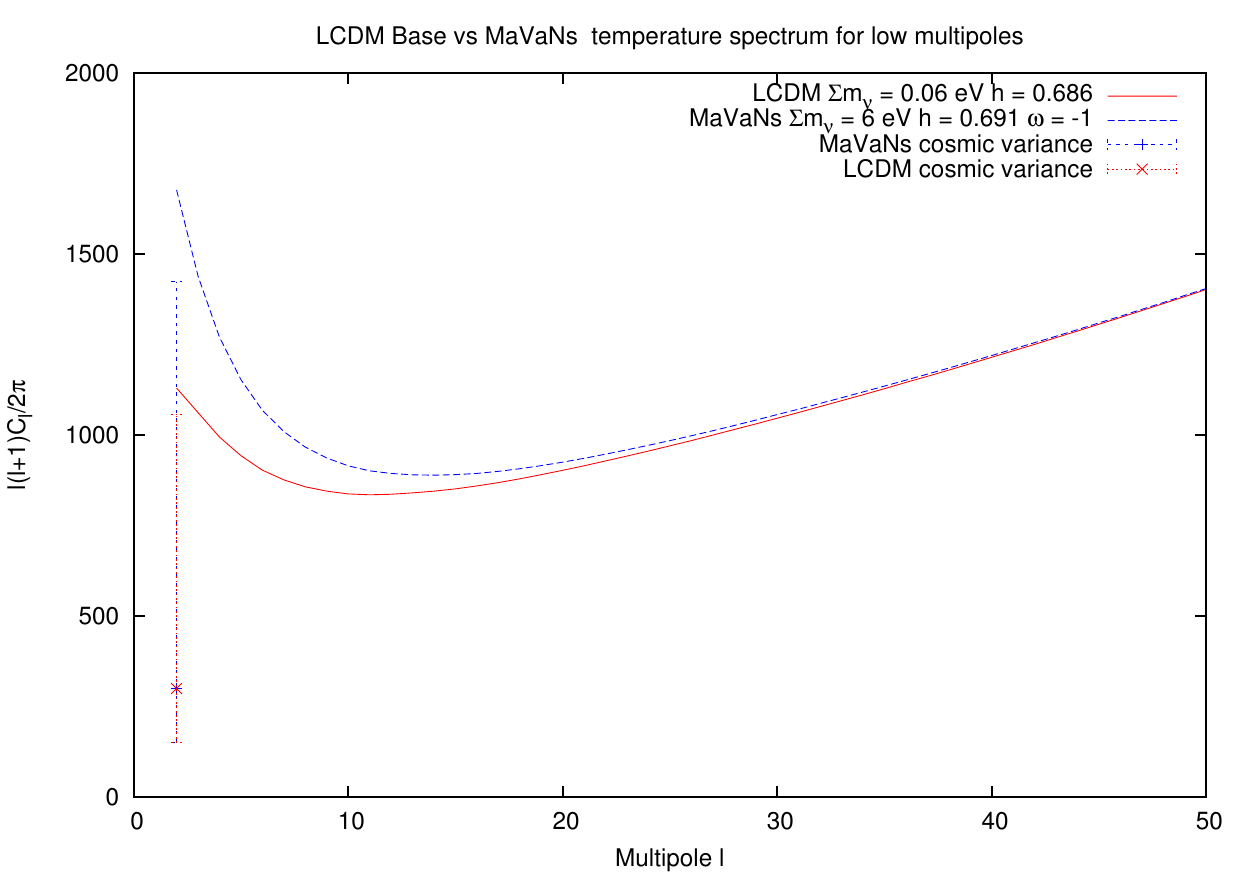}
\caption{\label{fig:epsart} This is same as figure 5 except for low $\it l$ modes. The red curve shows the best fit temperature spectrum for our base $\Lambda$CDM cosmology.The blue curve show the best fit temperature spectrum for a MaVaNs cosmology with $\omega$ = -1 and $\Sigma$m$_\nu$ = 6 eV. Their corresponding error bars for the quadrupole are also shown. The two curves are different for $\it l$ $\le$ 10 but start concurring as we go higher in $\it l$.}
\end{figure}

We have tabulated our results for the cosmological parameters of various cosmologies in table 1. Since we are scanning the parameter space over 4 parameters, the criterion for the error bars is that the log likelihood should not be more than 2.38 lower compared to the best fit model. This gives us our 68$\%$ bounds. $\Omega_b$h$^2$ and $\Omega_m$h$^2$ have not been listed because they were not found to change significantly with the model. This makes sense since the neutrinos in both cosmologies are effectively massless at and before recombination. This implies that recombination must have happened at the same redshift in either theory and hence $\rho_b$ and $\rho_m$ should be the same at recombination and hence throughout history in both cosmologies. This is also what we see in our likelihood fits. The only parameters that change significantly within cosmologies are H$_0$ and $\sigma_8$. H$_0$ is different between MaVaNs and base $\Lambda$CDM since the distance to last scattering is different in both theories for the same H$_0$ which will result in shifting in the position of the acoustic peaks. Thus H$_0$ has to be tuned in MaVaNs to get the correct distance to last scattering. The H$_0$  decreases with increasing $\omega$. For $\omega$ = -1, the best fit H$_0$ is slightly higher than the $\Lambda$CDM base model. For $\omega$ = -0.93 it is already significantly lower that $\Lambda$CDM base model. So as to not increase the tension between CMB measurements of H$_0$ and other measurements of H$_0$, $\omega$ lower than -0.93 is probably not feasible.

\subsection{Possible Application to the $\sigma_8$ discrepancy in Planck}

The Planck Collaboration recently reported their results of the measurement of the anisotropy of the CMB background \cite{PlanckCosmoParameters:2013} and found it to be consistent with $\Lambda$CDM Cosmology. They reported the measured value of the RMS fluctuations of matter density in linear theory today to be                                                       
\begin{equation}                                                                                                                    
\sigma_{8}=0.834\pm0.027 (68\%;Planck,\Sigma m_\nu=0.06 eV)                                                           
\end{equation}                                                                                                                      

One can also determine the RMS fluctuations of the matter density by measuring cluster counts as a function of redshift. The Planck Collaboration measured this function from the Sunyaev-Zeldovich (SZ) effect on the CMB photons along whose line of sight the clusters lie. They used a sample of 189 clusters for whom the signal-to-noise ratio was more than seven. They found that the number of clusters in each red shift bin was significantly smaller than what you would expect from $\Lambda$CDM cosmology. They measured the RMS 
value of the matter density fluctuations to be \cite{PlanckSZ:2013}                                                                      
                                           
\begin{equation}                                                                                                                   
\sigma_{8} = 0.77\pm0.02                                                                 
\end{equation}
in tension with the value derived from the temperature spectrum.                                                                    
                                          
One way to reduce this tension would be to have a higher neutrino mass. However in $\Lambda$CDM cosmology increasing the neutrino mass decreases significantly the value of Hubble's constant in order to get a good fit to the temperature power spectrum which increases the already existing tension between H$_0$ as measured by Planck and other experiments\cite{PlanckCosmoParameters:2013}.

In order to resolve this issue we need to be able to change the matter power spectrum without affecting the temperature power spectrum. MaVaNs are a candidate for this purpose since they act as effectively massless during recombination but become massive later acting as warm dark matter. As evidenced by table 1 using MaVaNs we can get a significant decrease in $\sigma_8$ as calculated in linear theory without changing $\it h$ significantly.

\begin{table*}                                                                                                                       
\caption{\label{tab:table1} Tabulated below are the values of cosmological parameters for different cosmologies. We do not quote $\Omega$$_m$h$^2$ and $\Omega$$_b$h$^2$, because they don't change significantly with the cosmology and have a best fit values around 0.0220 and 0.1385 respectively. The quantities in brackets give the best fit parameters.}                                                                                                           
\begin{ruledtabular}                                                                                                                
\begin{tabular}{cccccc}                                                                                                             
Cosmology & $\omega$ & $\Sigma$m$_\nu$ (eV) & $\it h$ & $\sigma_8$\\                                                                
\hline                                                                                                                              
$\Lambda$CDM & -1 & 0.06 &0.6855 $\pm$ 0.0125 (0.686) &0.824 $\pm$ 0.013 (0.824) \\                                                
MaVaNs       & -1 & 3.00 &0.691 $\pm$ 0.013 (0.691) &0.820 $\pm$ 0.013 (0.820)   \\                                           
MaVaNs       & -1 & 6.00 &0.691 $\pm$ 0.012 (0.691) &0.806 $\pm$ 0.012 (0.807)  \\
MaVaNs       & -0.96 & 3.00 &0.681 $\pm$ 0.012 (0.680) &0.8135 $\pm$ 0.0125 (0.814)  \\
MaVaNs       & -0.96 & 6.00 &0.678 $\pm$ 0.014 (0.680) &0.801 $\pm$ 0.013 (0.799)    \\
MaVaNs       & -0.93 & 3.00 &0.6715 $\pm$ 0.0125 (0.672) &0.8075 $\pm$ 0.0125 (0.807)  \\
MaVaNs       & -0.93 & 6.00 &0.669 $\pm$ 0.013 (0.672) &0.793 $\pm$ 0.013 (0.792)               \\                  
\end{tabular}                                                                                                                       
\end{ruledtabular}                                                                                                                  
\end{table*}

\section{Conclusions}
We have shown when the other cosmological parameters are allowed to vary , a good fit can be obtained to the   temperature fluctuation spectrum, even though the neutrino mass and hence the neutrino energy density become important at late times. The only cosmological parameters affected are H$_0$ (to match the distance to last scattering) and $\sigma_8$. We find that we can obtain a significantly smaller $\sigma_8$ in MaVaNs as compared to $\Lambda$CDM without changing  H$_0$ very much. Hence MaVaNs are a possible solution to Planck $\sigma_8$ discrepancy. Including the low $\it l$ data in the likelihood calculation will help us put upper bounds on the current neutrino mass. We leave this for future studies. 

 An interesting corollary to these results is that in MaVaNs theories, CMB data do not  constrain possible neutrino mass results in terrestrial experiments. A discrepancy between  the neutrino mass as determined from the CMB fits and the neutrino mass determined locally would be strong evidence for MaVaNs.
  
The $\sigma_8$ that we have calculated here has been done in linear theory. Structure formation simulations which include mass varying neutrinos are called for to establish these results conclusively. In our study we assumed that the neutrinos do not clump. One can also study the cosmology in the case of a heavier acceleron where neutrino clumping occurs and `neutrino nuggets' are formed. It will also be interesting to check the MaVaNs scenario described by \cite{Fardon:2006}. It also remains to be checked whether MaVaNs is consistent with other data sets such as BAO. We leave this for future studies.

\section{Acknowledgements}

The authors would like to thank the creators of CMBEASY code for making such a excellent and well documented code available to the public. We also would like to thank Tom Quinn for helpful discussions on the subject.   Partial support was provided by the DOE under Grant No. DE-FG02-96ER40956.  

\section{Appendix: List of nuisance parameters used}

The nuisance parameters labelled as Planck were obtained from Table 5 of \cite{PlanckCosmoParameters:2013} and are the best fit values for Planck+WP data set. The remaining ones were guessed based on the priors listed in Table 4 of the same paper.\\
$A^{PS}_{100}$ = 152 (Planck)\\ 
$A^{PS}_{143}$ = 63.3 (Planck)\\
$A^{PS}_{217}$ = 117.0 (Planck)\\
$A^{CIB}_{143}$ = 0.0 (Planck)\\
$A^{CIB}_{217}$ = 27.2 (Planck)\\
$A^{tSZ}_{143}$ = 6.80 (Planck)\\
$r^{PS}_{143*217}$ = 0.916 (Planck)\\
$r^{CIB}_{143*217}$ = 0.406 (Planck)\\
$\gamma$$^{CIB}$ = 0.601 (Planck)\\
$c_{100}$ = 1.0006\\
$c_{217}$ = 0.9966\\
$\xi$$^{tsz*CIB}$ = 0.03(Planck)\\
$A^{ksz}$ = 0.9(Planck)\\
$\beta$$^{i}$$_{j}$ = 1\\

\bibliographystyle{h-physrev}
\bibliography{neutrino}




   
    


\end{document}